\documentclass[showpacs,prl,twocolumn,aps,preprintnumbers]{revtex4}
\usepackage{amsmath,amssymb}

\newcommand{\beq}{\begin{equation}}
\newcommand{\eeq}{\end{equation}}
\newcommand{\ben}{\begin{eqnarray*}}
\newcommand{\een}{\end{eqnarray*}}

\newcommand\tr{\mathop{\mathrm{tr}}}

\begin{document}
 \preprint{BNL-94104-2010-JA}
 \preprint{INT-PUB-10-053}
\title{Testing the chiral magnetic and chiral vortical effects \\
in heavy ion collisions}

\author{Dmitri E. Kharzeev$^{1,2}$ and Dam T. Son$^3$}

\affiliation{$^1$Department of Physics and Astronomy, Stony Brook University, Stony Brook, New York 11794-3800, USA\\
$^2$Physics Department, Brookhaven National 
Laboratory, Upton, New York 11973-5000, USA\\
$^3$Institute for Nuclear Theory, University of Washington, Seattle, 
Washington 98195-1550, USA}

\begin{abstract}
We devise a test of the Chiral Magnetic and Chiral Vortical effects
(CME and CVE) in relativistic heavy ion collisions that relies only on
the general properties of triangle anomalies. We show that the ratio
$R_{EB}=J_E/J_B$ of charge $J_E$ and baryon $J_B$ currents for CME is
$R^{\rm CME}_{EB} \to \infty$ for three light flavors of quarks ($N_f
=3$), and $R^{\rm CME}_{EB} = 5$ for $N_f =2$, whereas for CVE it is
$R^{\rm CVE}_{EB} =0$ for $N_f =3$ and $R^{\rm CME}_{EB} = 1/2$ for
$N_f =2$. The physical world with light $u,d$ quarks and a heavier $s$
quark is in between the $N_f =2$ and $N_f =3$ cases; therefore, the
ratios $R_{EB}$ for CME and CVE should differ by over an order of
magnitude.  Since the ratio of electric charge and baryon asymmetries
is proportional to $R_{EB}$, the measurement of baryon and electric
charge asymmetry fluctuations should allow to separate clearly the CME
and CVE contributions. In both cases, there has to be a positive correlation between the charge 
and baryon number asymmetries that can be tested on the event-by-event basis. At a lower collision energy, as the baryon number
density increases and the CVE potentially plays a r\^ole, we expect
the emergence of the baryon number asymmetry.
\end{abstract}

\pacs{11.40.Ha,12.38.Mh,25.75.Ag}
\maketitle


Recently, STAR \cite{:2009uh, :2009txa} and PHENIX \cite{phenix}
Collaborations at Relativistic Heavy Ion Collider reported
experimental observation of charge asymmetry fluctuations.  While the
interpretation of the observed effect is still under intense
discussion, the fluctuations in charge asymmetry have been predicted
to occur in heavy ion collisions due to the Chiral Magnetic Effect
(CME) in QCD coupled to electromagnetism
\cite{Kharzeev:2004ey,Kharzeev:2007tn,Kharzeev:2007jp,Fukushima:2008xe,Kharzeev:2009fn}.
Related phenomena have been discussed in the physics of
primordial electroweak plasma \cite{Giovannini:1997gp} and quantum
wires \cite{acf}.  The CME scenario assumes a chirality asymmetry
between left- and right-handed quarks, parametrized by an axial
chemical potential $\mu_5$.  Such an asymmetry can arise if there is
an asymmetry between the instanton and anti-instanton events (or any
topology-changing transitions in general) early in the heavy ion
collision.  The chirality asymmetry, coupled to a strong magnetic
field created by the colliding ions
\cite{Kharzeev:2007jp,Skokov:2009qp}, generates a current of electric
charge.  This is the CME, which is one of several effects arising from triangle
anomalies in the medium.  

A related effect---the emergence of a chiral current in a medium with
finite baryon density, in an external magnetic field or in the
presence of a vorticity the fluid---has also been discussed in the
literature~\cite{Son:2004tq,Metlitski:2005pr,Son:2009tf}.
The close connection between CME and
the latter effect can be established for example by the method of dimensional
reduction appropriate in the case of a strong magnetic field
\cite{Basar:2010zd}: the simple relations $J_V^0 = J_A^1, \ J_A^0 =
J_V^1$ between the vector $J_V$ and axial $J_A$ currents in the
dimensionally reduced $(1+1)$ theory imply that the density of
baryon charge must induce the axial current, and the density of
axial charge must induce the charge current (CME).

The CME can be derived in several ways.  A heuristic explanation is as
follows: magnetic field leads to the spin polarization of quarks.  But
since there are, say, more right-handed quarks than left-handed
quarks, the quarks will preferably move along the direction of the
magnetic field, leading to a current.  More rigorously, if one solves
the Dirac equation in external magnetic field, one finds that the
lowest Landau level is chiral.  When there is a chemical potential for
the axial charge, some of the energy levels in the lowest Landau
levels are filled, inducing a nonzero current.

One may worry that the single-particle picture based on the Dirac
equation will cease working once an interaction is turned on.  However,
the essentially topological nature of the phenomenon guarantees the result
even in the presence of interaction.  In particular, in holographic
models (at infinite 't Hooft coupling) the magnitude of the chiral
magnetic effect \cite{Yee:2009vw,Rebhan:2009vc,Gorsky:2010xu} appears the
same as at weak coupling \cite{Rubakov:2010qi,Gynther:2010ed,Yee:2009vw}.  The
CME has been studied in lattice QCD coupled to electromagnetism, both in the quenched
\cite{Buividovich:2009wi,Buividovich:2009zzb,Buividovich:2010tn} and
dynamical (domain wall) fermion \cite{Abramczyk:2009gb} formulations.

It is important to establish whether the CME explanation of charge
asymmetry fluctuations is the correct one.  First, it would be a
direct observation of a topological effect in QCD.  Second, the
magnitude of this effect in the chirally broken phase is expected to
be much smaller and hence the observation of the CME would manifest the restoration of 
chiral symmetry in the medium.  The effort of
quantifying the charge asymmetry fluctuations in QCD matter and of
examining alternative explanations and backgrounds has already begun
(see
e.g. \cite{Bzdak:2009fc,Nam:2009jb,Fukushima:2010vw,Gorsky:2010dr,Fu:2010rs,Schlichting:2010na,Asakawa:2010bu,Voloshin:2010ut,Orlovsky:2010ga,Zhitnitsky:2010zx,Muller:2010jd,Mages:2010bc}; ref. \cite{Rogachevsky:2010ys} discusses also the baryon asymmetries), 
and there are plans to further study this effect at RHIC, LHC, FAIR
and NICA.

\vskip0.3cm
Due to the importance of the question, we need to devise tests for the
CME mechanism.  In this letter we propose such a test.  Our
proposal relies on two recent findings.  The first is that the matter
created at RHIC behaves as an almost perfect liquid: hydrodynamic
models describe the gross properties of the droplet very well (for review, see \cite{Schafer:2009dj}).  The
second finding is that quantum anomalies modify the hydrodynamics of a
relativistic fluid.  In addition to the chiral magnetic effect, there
is also a chiral vortical effect: the vorticity $\vec\omega$, combined
with a baryon chemical potential $\mu_B$, creates an effective
magnetic field $\mu_B\vec\omega$.  Therefore one has, in addition to
the CME, a Chiral Vortical Effect (CVE).  The exact magnitude of the effect in relativistic 
hydrodynamics has been found in Ref.~\cite{Son:2009tf}, but
its existence has been proposed before~\cite{Kharzeev:2007tn}.
Vorticity in heavy ion collisions
is a natural consequence of the angular momentum conservation (see
e.g. \cite{Kharzeev:2004ey,Liang:2004ph,Betz:2007kg,Becattini:2007zn});
the estimates of vorticity and the discussion of its r{\^o}le in heavy
ion collisions can be found in \cite{Becattini:2007sr}.


Let us first recall the general formulae for anomalous
hydrodynamics~\cite{Son:2009tf}.  Suppose that the system under
consideration has a chemical potential $\mu$, coupled to a charge
$\bar q\gamma^0 B q$, where $B$ is a flavor matrix, and an axial
chemical potential $\mu_5$, coupled to the axial charge $\bar q
\gamma^0\gamma^5 A q$, where $A$ is another flavor matrix.  For
simplicity, we shall assume that both $\mu$ and $\mu_5$ are much
smaller than the temperature $T$ (this assumption usually holds in relativistic heavy ion collisions).  We also assume that electromagnetism
couples to the current $\bar q \gamma^\mu Q q$, with $Q$ being the
charge matrix.  If one measures a vector current $J^\mu = \bar
q\gamma^\mu V q$, then the result is
\begin{equation}
  \vec J = \frac{N_c\mu_5}{2\pi^2} [\tr (V\!AQ) \vec B + \tr(V\!AB)2\mu \vec \omega]
\end{equation}
where $\vec B$ and $\vec \omega$ are the external magnetic fields and
the fluid vorticity respectively.  The two parts of the current on the right hand
side correspond to the CME and the CVE, respectively.  The traces in
the formula are related to the anomalous triangle diagram.

We shall consider two cases: $N_f=3$, where $u$, $d$ and $s$ quarks
are light, and $N_f=2$ where only $u$ and $d$ quarks are light.  In
both cases, we assume $A$ to be the unity matrix, $A=\openone$ (which
is expected if the chiral asymmetry is due to instanton events, which
are flavor symmetric), and $B=(1/3)\openone$.  For $N_f=3$, $Q={\rm
diag}(2/3,-1/3,-1/3)$, and for $N_f=2$, $Q={\rm diag}(2/3,-1/3)$.
There are two currents that we will measure: the electromagnetic
current $J_E$, corresponding to $V=Q$ and the baryon current $J_B$,
corresponding to $V=B$.

For CME, we get for the charge current (up to an overall factor of
$ N_c\ \mu_5 \vec B/(2\pi^2)$ which is common for both charge and baryon currents)
\beq
J_E^{CME} \sim \frac{2}{3} \ \  (N_f=3)\ \   {\rm or} \ \ \ \frac{5}{9} \ \ (N_f=2)
\eeq
and for the baryon current
\beq
J_B^{CME} = 0 \ \  (N_f=3)\ \   {\rm or} \ \ \ \sim \frac{1}{9} \ \ (N_f=2).
\eeq
For CVE, the results are (up to the overall factor $N_c\ \mu_5\mu \vec\omega/\pi^2$)
\beq
J_E^{CVE} = 0  \ \  (N_f=3)\ \   {\rm or} \ \ \ \sim \frac{1}{3} \ \ (N_f=2);
\eeq
\beq
J_B^{CVE} \sim 1 \ \  (N_f=3)\ \   {\rm or} \ \ \ \sim \frac{2}{3} \ \ (N_f=2).
\eeq

In the SU(3) case, the CME and CVE lead to completely different
currents: the CME contributes only to the electromagnetic current and the
CVE contributes only to the baryon current.  In the SU(2) case, the
separation is less clean, but the ratio of $J_B/J_E$ still differs by a
factor of ten.

Let us now discuss the implications of our calculation in heavy ion
collisions.  It is known that the baryon chemical potential of the produced fireball
depends on the collision energy: at smaller $\sqrt s$, $\mu$ is
larger.  Thus the CVE should be more important at lower energies.  According
to the computation above, $J_B/J_E$ becomes larger as one lowers the
energy of the collision.  Moreover, since the symmetry arguments suggest that the magnetic
field and the vorticity of the fluid have to be aligned, our results show that the two vectors
$\vec J_B$ and $\vec J_E$ should point in the same direction.

We can now formulate our predictions.  In addition to the charge separation,
there must be a baryon number separation.  The two effects are
positively correlated on the event-by-event basis, and the relative
importance of baryon number separation increases as one lowers the
energy of the collision.
Our predictions can be summarized as follows:
\begin{itemize}
\item There should be a baryon number separation of the same sign as the
electric charge separation;
\item The ratio between the baryon asymmetry and charge asymmetry should increase as
the center of mass energy is lowered; 
\item The magnitude of the ratio of charge and baryon asymmetries allows to discriminate between the CME and CVE mechanisms.

\end{itemize}
As our calculation depends on very few assumptions about the properties of the quark
gluon plasma beside the existence of the initial chirality imbalance,
the predictions above can be viewed as a nontrivial test for the CME
explanation of the charge asymmetry fluctuations at RHIC.  A priori,
the charge asymmetry and baryon asymmetry do not have to be
correlated, but this correlation must exist if CME is the mechanism
underlying charge asymmetry fluctuations.

Let us discuss the uncertainties of our predictions. Our computation of the ratio of charge and baryon currents 
depends only on the general properties of the triangle anomalies, and so should be robust. The  
experimental study of baryon asymmetry would ideally require the measurement of all produced 
baryons and anti-baryons within a certain (symmetric) rapidity interval. Since this may not be feasible, our prediction 
would have to be supplemented by an assumption about the relative contributions of protons, neutrons and hyperons (and the 
corresponding anti-baryons).   It is highly desirable to perform a reliable evaluation of the absolute values of currents and 
asymmetries, and not just of their ratios. This computation would require a quantitative control over the magnitude of vorticity in the produced 
quark-gluon fluid and a treatment of the time evolution of vorticity and of magnetic field; such a study could be performed by methods of relativistic magnetohydrodynamics taking account of triangle anomalies. 

\vskip0.3cm
To summarize, we propose to test the CME and CVE in heavy ion collisions by the event-by-event study of correlations between the electric charge and baryon number asymmetries. We have evaluated the ratios of electric charge and baryon number asymmetries for CME and CVE mechanisms; our calculations depend only on the general properties of triangle anomalies.

We thank Valery Rubakov for discussions. The work of DK was supported in part by the U.S. Department of Energy 
under Contract No.~DE-AC02-98CH10886.  The work of DTS was supported in part 
by the U.S.\  Department of Energy 
under Contract No.~DE-FG02-00ER41132.


\begin{thebibliography}{00}
\bibitem{:2009uh}
  B.~I.~Abelev {\it et al.}  [STAR Collaboration],
  Phys.\ Rev.\ Lett.\  {\bf 103}, 251601 (2009)
  [arXiv:0909.1739 [nucl-ex]].

\bibitem{:2009txa}
  B.~I.~Abelev {\it et al.}  [STAR Collaboration],
  arXiv:0909.1717 [nucl-ex].

\bibitem{phenix}
A. Ajitanand, S. Esumi, R. Lacey [PHENIX Collaboration], in: Proc. of the RBRC Workshops, vol. 96, 2010:  
``P- and CP-odd effects in hot and dense matter''; http://quark.phy.bnl.gov/~kharzeev/cpodd/

\bibitem{Kharzeev:2004ey}
  D.~Kharzeev,
  Phys.\ Lett.\  B {\bf 633}, 260 (2006)
  [arXiv:hep-ph/0406125].

\bibitem{Kharzeev:2007tn}
  D.~Kharzeev and A.~Zhitnitsky,
  Nucl.\ Phys.\  A {\bf 797}, 67 (2007)
  [arXiv:0706.1026 [hep-ph]].


\bibitem{Kharzeev:2007jp}
  D.~E.~Kharzeev, L.~D.~McLerran and H~.J.~Warringa,
  Nucl. Phys.  A {\bf 803}, 227 (2008);

\bibitem{Fukushima:2008xe}
  K.~Fukushima, D.~E.~Kharzeev and H.~J.~Warringa,
  Phys.\ Rev.\  D {\bf 78}, 074033 (2008)
  [arXiv:0808.3382 [hep-ph]].
  
  \bibitem{Kharzeev:2009fn}
  D.~E.~Kharzeev,
  Annals Phys.\  {\bf 325}, 205 (2010)
  [arXiv:0911.3715 [hep-ph]].
  
\bibitem{Giovannini:1997gp}
  M.~Giovannini and M.~E.~Shaposhnikov,
  Phys.\ Rev.\ Lett.\  {\bf 80}, 22 (1998)
  [arXiv:hep-ph/9708303]; 
  Phys.\ Rev.\  D {\bf 57}, 2186 (1998)
  [arXiv:hep-ph/9710234].
   
  \bibitem{acf}
  A.Yu. Alekseev, V.V. Cheianov, J. Fr{\"o}lich, Phys.\ Rev.\ Lett. {\bf 81}, 3503 (1998).
  
\bibitem{Skokov:2009qp}
  V.~Skokov, A.~Y.~Illarionov and V.~Toneev,
  Int.\ J.\ Mod.\ Phys.\  A {\bf 24}, 5925 (2009)
  [arXiv:0907.1396 [nucl-th]].

\bibitem{Son:2004tq}
  D.~T.~Son and A.~R.~Zhitnitsky,
  Phys.\ Rev.\  D {\bf 70}, 074018 (2004)
  [arXiv:hep-ph/0405216].
  
\bibitem{Metlitski:2005pr}
  M.~A.~Metlitski and A.~R.~Zhitnitsky,
  Phys.\ Rev.\  D {\bf 72}, 045011 (2005)
  [arXiv:hep-ph/0505072].


\bibitem{Son:2009tf}
  D.~T.~Son and P.~Sur\'owka,
  Phys.\ Rev.\ Lett.\  {\bf 103}, 191601 (2009)
  [arXiv:0906.5044 [hep-th]].
  
\bibitem{Basar:2010zd}
  G.~Ba\c sar, G.~V.~Dunne and D.~E.~Kharzeev,
  Phys.\ Rev.\ Lett.\  {\bf 104}, 232301 (2010)
  [arXiv:1003.3464 [hep-ph]].

  
\bibitem{Yee:2009vw}
  H.~U.~Yee,
  JHEP {\bf 0911}, 085 (2009)
  [arXiv:0908.4189 [hep-th]]; K.~Y.~Kim, B.~Sahoo and H.~U.~Yee,
  arXiv:1007.1985 [hep-th].
  
\bibitem{Rebhan:2009vc}
  A.~Rebhan, A.~Schmitt and S.~A.~Stricker,
  JHEP {\bf 1001}, 026 (2010)
  [arXiv:0909.4782 [hep-th]].
  
\bibitem{Gorsky:2010xu}
  A.~Gorsky, P.~N.~Kopnin and A.~V.~Zayakin,
  arXiv:1003.2293 [hep-ph].
  
\bibitem{Rubakov:2010qi}
  V.~A.~Rubakov,
  arXiv:1005.1888 [hep-ph].
  
\bibitem{Gynther:2010ed}
  A.~Gynther, K.~Landsteiner, F.~Pena-Benitez and A.~Rebhan,
  arXiv:1005.2587 [hep-th].

\bibitem{Buividovich:2009wi}
  P.~V.~Buividovich, M.~N.~Chernodub, E.~V.~Luschevskaya and M.~I.~Polikarpov,
  Phys.\ Rev.\  D {\bf 80}, 054503 (2009)
  [arXiv:0907.0494 [hep-lat]].
  
\bibitem{Buividovich:2009zzb}
  P.~V.~Buividovich, E.~V.~Luschevskaya, M.~I.~Polikarpov and M.~N.~Chernodub,
  JETP Lett.\  {\bf 90}, 412 (2009)
  [Pisma Zh.\ Eksp.\ Teor.\ Fiz.\  {\bf 90}, 456 (2009)].
  
\bibitem{Buividovich:2010tn}
  P.~V.~Buividovich, M.~N.~Chernodub, D.~E.~Kharzeev, T.~Kalaydzhyan, E.~V.~Luschevskaya and M.~I.~Polikarpov,
  Phys.\ Rev.\ Lett.\  {\bf 105}, 132001 (2010)
  [arXiv:1003.2180 [hep-lat]].
  
\bibitem{Abramczyk:2009gb}
  M.~Abramczyk, T.~Blum, G.~Petropoulos and R.~Zhou,
  arXiv:0911.1348 [hep-lat].
  
\bibitem{Bzdak:2009fc}
  A.~Bzdak, V.~Koch and J.~Liao,
  Phys.\ Rev.\  C {\bf 81}, 031901 (2010)
  [arXiv:0912.5050 [nucl-th]]; arXiv:1005.5380 [nucl-th]; arXiv:1008.4919 [nucl-th].

\bibitem{Nam:2009jb}
  S.-i.~Nam,
  Phys.\ Rev.\  D {\bf 80}, 114025 (2009)
  [arXiv:0911.0509 [hep-ph]]; Phys.\ Rev.\  D {\bf 82}, 045017 (2010)
  [arXiv:1004.3444 [hep-ph]].


\bibitem{Fukushima:2010vw}
  K.~Fukushima, D.~E.~Kharzeev and H.~J.~Warringa,
  Phys.\ Rev.\ Lett.\  {\bf 104}, 212001 (2010)
  [arXiv:1002.2495 [hep-ph]]; Nucl.\ Phys.\  A {\bf 836}, 311 (2010)
  [arXiv:0912.2961 [hep-ph]].

\bibitem{Gorsky:2010dr}
  A.~Gorsky and M.~B.~Voloshin,
  arXiv:1006.5423 [hep-th].

\bibitem{Fu:2010rs}
  W.-j.~Fu, Y.-x.~Liu and Y.-l.~Wu,
  arXiv:1003.4169 [hep-ph].

\bibitem{Schlichting:2010na}
  S.~Schlichting and S.~Pratt,
  arXiv:1005.5341 [nucl-th]; arXiv:1009.4283 [nucl-th].
  
\bibitem{Asakawa:2010bu}
  M.~Asakawa, A.~Majumder and B.~M\"uller,
  Phys.\ Rev.\  C {\bf 81}, 064912 (2010)
  [arXiv:1003.2436 [hep-ph]].
  

\bibitem{Voloshin:2010ut}
  S.~A.~Voloshin,
  arXiv:1006.1020 [nucl-th].
  
\bibitem{Orlovsky:2010ga}
  V.~Orlovsky and V.~Shevchenko,
  arXiv:1008.4977 [hep-ph].
  
\bibitem{Zhitnitsky:2010zx}
  A.~R.~Zhitnitsky,
  arXiv:1008.3598 [nucl-th].
  
\bibitem{Muller:2010jd}
  B.~M\"uller and A.~Sch\"afer,
  arXiv:1009.1053 [hep-ph].
  
\bibitem{Mages:2010bc}
  S.~W.~Mages, M.~Aicher and A.~Sch\"afer,
  arXiv:1009.1495 [hep-ph].
  
\bibitem{Rogachevsky:2010ys}
  O.~Rogachevsky, A.~Sorin and O.~Teryaev,
  arXiv:1006.1331 [hep-ph].
  
\bibitem{Schafer:2009dj}
  T.~Sch\"afer and D.~Teaney,
  Rept.\ Prog.\ Phys.\  {\bf 72}, 126001 (2009)
  [arXiv:0904.3107 [hep-ph]].
   
\bibitem{Liang:2004ph}
  Z.~T.~Liang and X.~N.~Wang,
  Phys.\ Rev.\ Lett.\  {\bf 94}, 102301 (2005)
  [Erratum-ibid.\  {\bf 96}, 039901 (2006)]
  [arXiv:nucl-th/0410079].
  
\bibitem{Betz:2007kg}
  B.~Betz, M.~Gyulassy and G.~Torrieri,
  Phys.\ Rev.\  C {\bf 76}, 044901 (2007)
  [arXiv:0708.0035 [nucl-th]].
  
\bibitem{Becattini:2007zn}
  F.~Becattini and L.~Ferroni,
  Eur.\ Phys.\ J.\  C {\bf 52}, 597 (2007)
  [arXiv:0707.0793 [nucl-th]].
  
\bibitem{Becattini:2007sr}
  F.~Becattini, F.~Piccinini and J.~Rizzo,
  Phys.\ Rev.\  C {\bf 77}, 024906 (2008)
  [arXiv:0711.1253 [nucl-th]].
\end{thebibliography}
\end{document}